\documentclass[aps,prl, showpacs, twocolumn, superscriptaddress]{revtex4-1}
\pdfoutput=1
\usepackage{graphicx}  
\usepackage{dcolumn}   
\usepackage{bm}        
\usepackage{amssymb}   
\usepackage{amsmath}   
\usepackage{dsfont} 
\usepackage[version=3]{mhchem} 

\newcommand{\ehpair}{$\emph{e}$-$\emph{h}$}
\newcommand{\eepair}{$\emph{e}$-$\emph{e}$}

\begin{document}
\title{Accurate X-Ray Absorption Predictions for Transition Metal Oxides: An Advanced Self-Consistent-Field Approach Inspired by Many-Body Perturbation Theory }

\author{Yufeng Liang}
\email{yufengliang@lbl.gov}
\affiliation{The Molecular Foundry, Lawrence Berkeley National Laboratory, Berkeley, CA 94720, USA}
\author{John Vinson}
\affiliation{National Institute of Standards and Technology (NIST), Gaithersburg, MD 20899, USA}
\author{Sri Pemmeraju}
\affiliation{The Molecular Foundry, Lawrence Berkeley National Laboratory, Berkeley, CA 94720, USA}
\author{Walter Drisdell}
\affiliation{Chemical Sciences Division, Lawrence Berkeley National Laboratory, Berkeley, CA 94720, USA}
\author{Eric Shirley}
\affiliation{National Institute of Standards and Technology (NIST), Gaithersburg, MD 20899, USA}
\author{David Prendergast}
\affiliation{The Molecular Foundry, Lawrence Berkeley National Laboratory, Berkeley, CA 94720, USA}

\begin{abstract}
Constrained-occupancy self-consistent-field ($\Delta$SCF) methods and many-body perturbation theories (MBPT) are two strategies for obtaining electronic excitations from first-principles. Using the two distinct approaches, we study the O $1s$ core excitations that have become increasingly important for characterizing transition metal oxides and developing theory of strong correlations. Interestingly, we find that the $\Delta$SCF approach, in its current single-particle form,  systematically underestimates the pre-edge intensity for chosen oxides, despite its success in weakly correlated systems. By contrast, the Bethe-Salpeter equation within MBPT predicts much better lineshapes. This inspires us to reexamine the many-electron dynamics of X-ray excitations. We find that the single-particle $\Delta$SCF approach can be rectified by explicitly calculating many-body transition amplitudes, producing X-ray spectra in excellent agreement with experiments. Our study paves the way to accurately predict X-ray near-edge spectral fingerprints for physics and materials science beyond the Bethe-Salpether equation.
\end{abstract}
\maketitle

X-ray absorption spectroscopy (XAS) is a powerful characterization technique to address challenging problems in physics, chemistry, and materials science, owing to its element specificity and orbital selectivity. 
With the help of density-functional theory (DFT), the interpretation of XAS is greatly facilitated by simulating spectral fingerprints for hypothetical structures from first-principles. 
Satisfactory X-ray absorption spectra have been simulated across a wide range of systems from small molecules \cite{uejio2008effects, triguero1998calculations} to condensed-matter systems \cite{mizoguchi2000core, mo2000ab, duscher2001core,  taillefumier2002x, hetenyi2004calculation, cavalleri2005half, prendergast2006x} and even complex interfaces \cite{velasco2014structure}.

Recently, interpreting XAS fingerprints for transition metal oxides (TMOs) from first-principles has become a pressing matter. 
In part, this is fueled by the quest for next-generation energy materials, for rechargeable battery cathodes \cite{yabuuchi2011detailed, hu2013origin, lin2016metal, luo2016charge}, fuel cells \cite{suntivich2011design, strasser2010lattice}, water-splitting catalysts \cite{zhu2012bonding, matsukawa2014enhancing, grimaud2013double}, and transparent conductive layers \cite{lebens2016direct}. Many prototypes of these energy devices are TMOs with complex chemical properties due to their $d$ orbitals.  
There is an enormous interest in how this electronic structure affects their functioning principles or failure mechanisms. 
Since the vast majority of these materials undergo inhomogeneous chemical reactions \cite{yabuuchi2011detailed, hu2013origin, lin2016metal}, and some of them exhibit complicated surface \cite{suntivich2011design, strasser2010lattice, zhu2012bonding, matsukawa2014enhancing} and interfacial \cite{grimaud2013double} behaviors, this makes X-ray spectroscopy particularly effective in revealing local chemical and structural properties.

Besides the immediate needs in materials research, understanding of the intriguing electron correlations inherent in TMOs can be advanced by the interpretation of core-level spectra. 
For instance, XAS has been employed to investigate the metal-insulator transitions induced by correlated effects \cite{koethe2006transfer, ruzmetov2007x, aetukuri2013control},  hole-doping-induced high-$T_c$ superconductivity \cite{chen1991electronic, peets2009x}, and, recently, emergent phenomena at perovskite interfaces \cite{lee2013titanium}. 
Since core-level spectra reflect the energy distribution of orbitals of particular symmetries, they often serve as powerful guides to advance theories for correlated electron systems, including the DFT+U method \cite{dudarev1998electron}, dynamical mean-field theory \cite{anisimov2009coulomb}, and exact diagonalization approaches \cite{wang2010theory, chen2013doping}. 

Of particular interest in the study of TMOs is the O $K$ edge that arises from the $1s\rightarrow 2p$ dipole-allowed transition. 
Since the \ce{O} $2p$ orbitals can hybridize covalently with the metal $d$ orbitals, the O $K$ edge contains information on the $d$-state electronic structure that is crucial to many interesting phenomena in TMOs. 
Unlike the transition metal $L$-edge ($2p\rightarrow d$) spectra, O $K$ edge spectra are not affected by core-level spin-orbit coupling and are free of atomic multiplet effects \cite{de2008core}. 
In almost all of the aforementioned examples \cite{yabuuchi2011detailed, suntivich2011design, lin2016metal, luo2016charge, zhu2012bonding, matsukawa2014enhancing, grimaud2013double, lebens2016direct, koethe2006transfer, ruzmetov2007x, chen1991electronic, peets2009x, lee2013titanium, dudarev1998electron,wang2010theory, chen2013doping}, there are measurements at the O $K$ edge. Despite its utility, very few studies have simulated this absorption edge for TMOs from first-principles, limiting the interpretation of the spectra, because of the difficulty in treating both strong electron-electron (\eepair{}) and electron-core-hole correlations adequately. 

In this Letter,  we present a first-principles study of the O $K$ edge of TMOs, using two state-of-the-art theories of XAS simulation: (1) the constrained-occupancy self-consistent field ($\Delta$SCF) core-hole approach \cite{taillefumier2002x, hetenyi2004calculation, prendergast2006x} and (2) the core-level Bethe-Salpeter equation (BSE) within many-body perturbation theory (MBPT) \cite{shirley1998ab, benedict1998optical, rehr2005final, vinson2011bethe}. 
Both theories yield absorbance $\sigma(\omega)$ with approximations to the many-body final state $|\Phi_f\rangle$ in Fermi's Golden rule,
\begin{equation}
\sigma(\omega)\propto \omega 
\sum_f |\langle\Phi_f|\bm{\epsilon}{\cdot}\bm{R}|\Phi_i\rangle|^2\delta(E_f-E_i-\hbar\omega)
\label{Fermi_golden}
\end{equation}
where $\bm{\epsilon}$ and $\bm{R}$ are the photon polarization and many-body position operator respectively, and $|\Phi_i\rangle$ is the initial state.
But the two theories differ in the underlying philosophy of solving the many-electron problem. 
As a $\Delta$SCF method, the core-hole approach iterates the total electron density self-consistently with \eepair{} interactions, while the BSE considers elementary excitations and treats \eepair{} interactions perturbatively. 
One would expect a self-consistent theory works better than a perturbation theory. We find, however, that the one-body core-hole approach fails to describe the pre-edge region of the O $K$ edge for TMOs, despite its success in producing XAS for many weakly correlated systems \cite{pascal2014x, mcdonald2015cooperative,pascal2014finite, prendergast2006x,velasco2014structure}. 
On the other hand, we show that the BSE is much more reliable for predicting XAS fingerprints for TMOs. The success of the BSE inspires us to examine both theories in a more general formalism, and ultimately demonstrate that the failure of the current $\Delta$SCF approach can be rectified through proper evaluation of the many-body matrix elements.

\begin{figure}[t]
  \centering
  \includegraphics[width=0.8\linewidth]{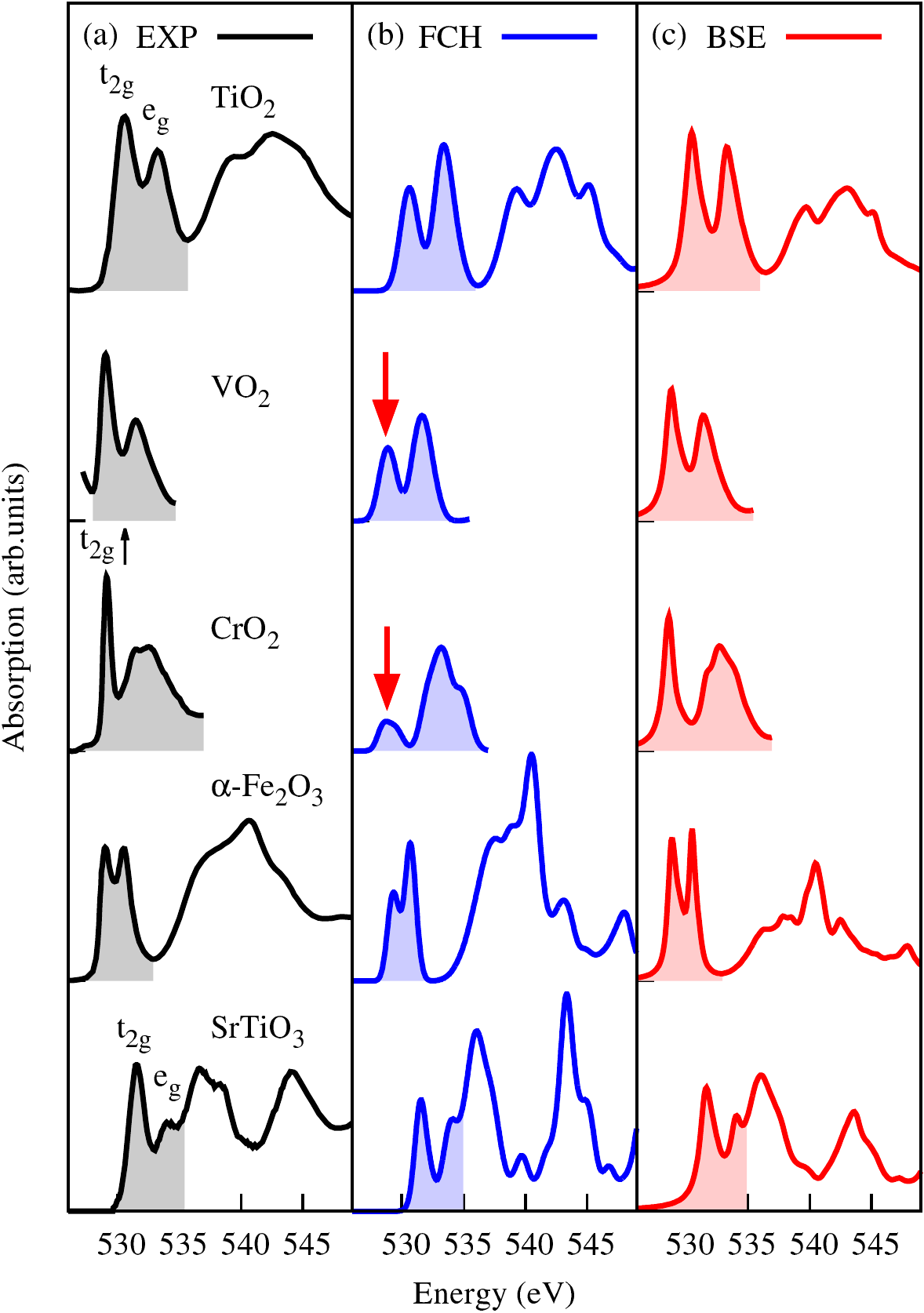}
  \caption{A comparison of experimental O $K$ edge (a) with the simulated spectra by the FCH approach (b) and BSE (c). The pre-edge regions are covered by shaded areas. Spectra are normalized according to the $e_g$ peak intensity. The severely underestimated $t_{2g}$ peaks are marked by red arrows. }
  \label{fig:spectra}
\end{figure}

Five TMOs are selected for benchmarking the first-principles XAS theories: the rutile phase of \ce{TiO2}, \ce{VO2}, and \ce{CrO2} as well as the the corundum $\alpha$-\ce{Fe2O3} and the perovskite \ce{SrTiO3}. They vary greatly in structure, band gap, or magnetism. The rutile \ce{VO2} ($> 340K$) and \ce{CrO2} are metallic, whereas \ce{TiO2} and \ce{SrTiO3} is insulating. \ce{CrO2} is ferromagnetic (FM) while \ce{Fe2O3} is antiferromagnetic (AFM). The rest exhibit no magnetism. The O $K$ edges from previous experiments \cite{yan2009oxygen, koethe2006transfer, stagarescu2000orbital, shen2014surface, zhu2012bonding} are shown in Fig. \ref{fig:spectra} (a). These spectra are angularly averaged except for \ce{CrO2}, where the polarization is perpendicular to the magnetization axis \cite{stagarescu2000orbital}. The \ce{TM}-$3d$-\ce{O}-$2p$ hybridization manifests as sharp double peaks around 530 eV, which result from the $t_{2g}$-$e_g$ splitting in the octahedral field. The intensity ratio of the two peaks is sensitive to changes in hybridization or charge transfer and often serves as a diagnostic tool.

\paragraph{Core-Hole Approach} 
The core-excited atom is treated as a single impurity with one electron removed from the excited core level. 
Depending on whether or not the X-ray photoelectron is added, the $\Delta$SCF core-hole approach is termed as an excited-electron and core-hole (XCH) or full core-hole (FCH) calculation \cite{taillefumier2002x, hetenyi2004calculation, prendergast2006x}. One then places the excited atom in a sufficiently large supercell to reduce spurious interactions between periodic replicas. The constrained-occupancy electron density is found using DFT. This modified ground state, including the presence of a core hole, is referred to as the \emph{final state}\cite{rehr2005final}, and that of the pristine system as the \emph{initial state}. 
The working approximation is to map the many-electron system onto a set of Kohn-Sham orbitals
\begin{equation}
\langle\Phi_f| \bm{\epsilon}{\cdot}\bm{R} |\Phi_i\rangle \approx  S\langle\tilde{\phi_f}| \bm{\epsilon}{\cdot}\bm{r} |\phi_h\rangle
\label{eq:matrix_element}
\end{equation}
where $\tilde{\phi_f}$'s are the unoccupied orbitals in the final state (with tilde) and $\phi_h$ is the core orbital in the initial state. $S$ is the many-body overlap and is usually treated as a constant independent of excitation energy.

To account for strong electron correlations in TMOs,  the DFT+U theory \cite{dudarev1998electron} is employed where the DFT energy is captured by the Perdew-Burke-Ernzerhof (PBE) functional, and the on-site Coulomb potential $U$ are from Ref.  \cite{wang2006oxidation}. We interpret the DFT+U orbital energies as quasiparticle (QP) energies and perform FCH rather than XCH calculations so as not to favor any particular orbital. More numerical details can be found in the Supplemental Material \cite{sm}.

Strikingly, the FCH approach systematically underestimates the near-edge peak associated with the $t_{2g}$ manifold for all selected TMOs. 
The $t_{2g}$ peak at $529.2$eV of \ce{CrO2} \cite{stagarescu2000orbital} suffers from the most severe underestimation. 
It becomes a weak, broad feature in the FCH simulation as compared to the strong, sharp peak in experiment. 
The $t_{2g}$-$e_g$ peak intensity ratios of \ce{VO2} and \ce{Fe2O3} are also too low - both are predicted as $0.7$, compared with $1.7$ and $1.0$ as measured, respectively. 
To validate the calculations, we have tested the numerical convergence with energy cutoffs and supercell sizes, and switched to local-density approximation functionals, but none of these suggests the failure of the FCH calculation is a numerical artifact or the impact of the empirical $U$ values \cite{sm}.

\paragraph{Bethe-Salpeter Equation} 
Within the Tamm-Dancoff approximation (TDA) \cite{benedict1998optical, rohlfing2000electron, onida2002electronic}, the photo-excited state is a superposition of electron-hole (\ehpair{}) pairs
\begin{equation}
|\Phi_f\rangle=\sum_{c}A^f_{c}a^\dagger_{c}h^\dagger|\Phi_i\rangle
\label{eq:tda}
\end{equation}
where $a^\dagger_{c}$ and $h^\dagger$ are the initial-state electron and the $1s$ core-hole creation operators respectively. The exciton amplitude $A^f_{c}$ can be solved from the BSE
\begin{equation}
(\varepsilon_c-\varepsilon_h)A^f_{c}+\sum_{c'}K^{eh}_{c, c'}A^f_{c'}=E_f A^f_{c}
\label{eq:bse}
\end{equation}
where $\varepsilon_{c,h}$ are single-particle energies of the initial-state orbitals. $K^{eh}$ is the \ehpair{} interaction kernel, comprising a direct term $K^{eh, \text{D}}$ and an exchange term $K^{eh, \text{X}}$. $E_f$ is the eigenenergy of the excitation. $K^{eh, \text{D}}$ involves the screened Coulomb interaction $W$ obtained within the random phase approximation (RPA), whereas $K^{eh, \text{X}}$ involves the bare Coulomb interaction. The core-level BSE calculation is done with the OCEAN code \cite{vinson2011bethe, gilmore2015efficient}. 

As is shown in Fig. \ref{fig:spectra} (c), the BSE substantially improves on the O $K$ edge lineshape. The edge-sharpness of the lower-energy peak is retrieved for all the investigated TMOs, particularly for \ce{CrO2}. The simulated $t_{2g}$-$e_g$ intensity ratios are almost as measured for \ce{TiO2}, $\alpha$-\ce{Fe2O3}, and \ce{SrTiO3} (below 537 eV). The BSE calculations systematically improve on the initial-state prediction of lineshape for the rutile group from previous studies \cite{de2008core, de2001high}.


\begin{figure}[t]
  \centering
  \includegraphics[width=0.9\linewidth]{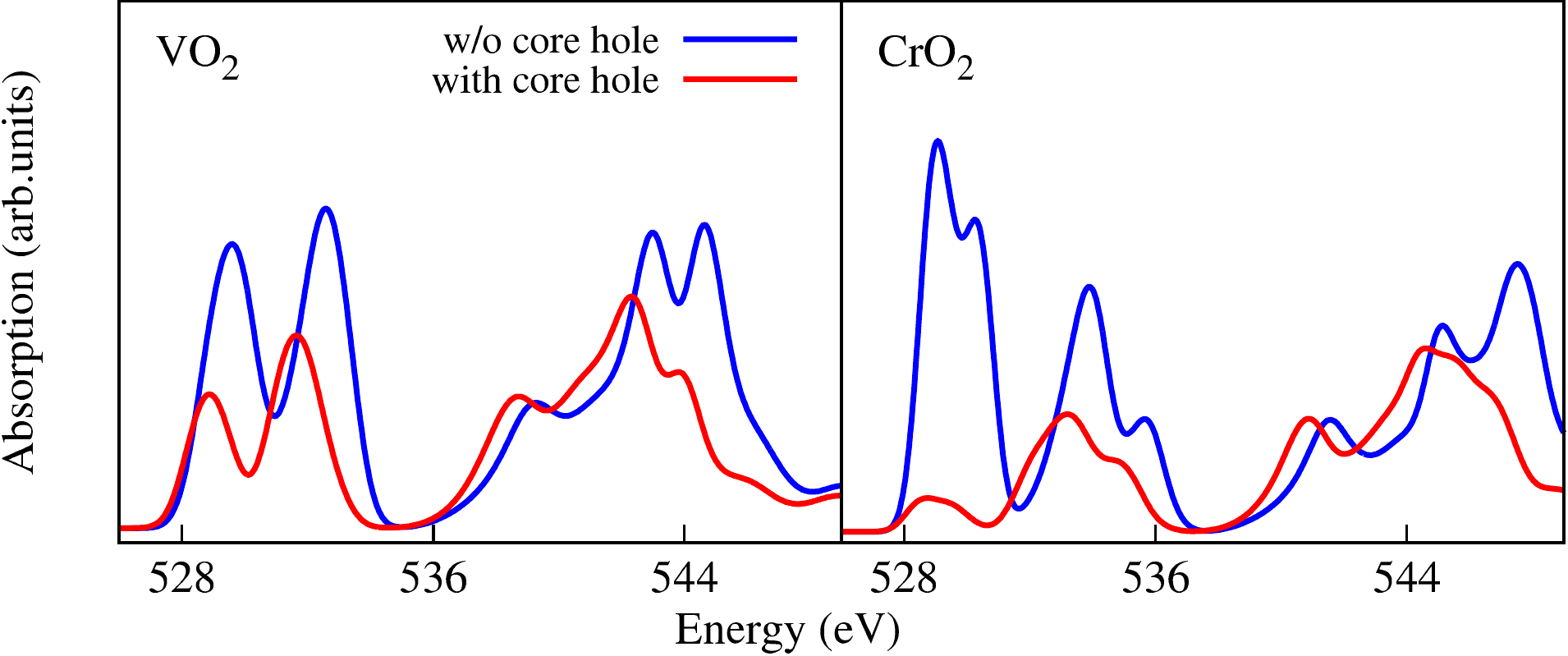}
  \caption{A comparison of the spectra with and without (w/o) the core hole in the FCH calculations. }
  \label{fig:compare}
\end{figure}

\paragraph{Discussion}
The core-hole effects in the O $K$ edge as predicted by the $\Delta$SCF method are counterintuitive. Typically, excitonic effects tend to sharpen the absorption edge due to \ehpair{} attraction \cite{benedict1998optical, rohlfing2000electron, shirley1998ab, duscher2001core}. In Fig. \ref{fig:compare} (b), we show the core-hole effects from FCH by comparing the initial- and final-state spectra. The core-hole attraction does redshift the spectra by more than $1$ eV in both the pre-edge and $4sp$  \cite{de1989oxygen} region (near $544$ eV). However, the FCH core-hole effect substantially reduces the $t_{2g}$ peak intensity.
Similar underestimated pre-peak intensity was encountered before  \cite{juhin2010angular, tanaka2005xanes, kanchana2006calculated} but no satisfactory explanation has been provided to date.

The failure of the $\Delta$SCF method in the pre-edge region motivates us to reexamine the approximations used in this approach as opposed to those in the BSE. 
The most obvious assumption that has been made is the approximation of the many-body matrix element in a single-body form. 
We now revisit the full many-body matrix element $\langle\Phi_f| \bm{\epsilon}{\cdot}\bm{R} |\Phi_i\rangle$. 
To this end, we express the final state $|\Phi_f\rangle$ in terms of the initial-state configurations. 
Since there is no restriction imposed on the initial configurations used in $\Delta$SCF, its adiabatic final state goes beyond TDA-BSE [Eq. (\ref{eq:tda})] and has incorporated the electronic response to the core-hole potential to all orders:
\begin{align}
\begin{split}
|\Phi_f\rangle=\big[\sum_{c}A^f_{c}a^\dagger_c h^\dagger + \sum_{c'cv}B^f_{c'cv}(a^\dagger_{c'} b^\dagger_{v}) a^\dagger_c h^\dagger 
+ \cdots\big]
|\Phi_i\rangle
\end{split}
\label{eq:Phi_f}
\end{align}
where $b_v^\dagger$'s are valence-hole creation operators. $c$($c'$) and $v$ sum over all the \emph{initial-state} empty and occupied orbitals respectively. 
However, despite the possible presence of one or more additional valence \ehpair{} pairs, such as $(a^\dagger_{c'} b^\dagger_{v}) a^\dagger_c h^\dagger$, only the leading-order coefficient $A^f_c$ can contribute to the many-body matrix element: $\langle\Phi_f| \bm{\epsilon}{\cdot}\bm{R} |\Phi_i\rangle=\sum_c(A^{f}_c)^* \langle \phi_c|\bm{\epsilon}\cdot\bm{r}|\phi_h\rangle$, because the dipole is a one-body operator.
Thus it is sufficient to find $A^f_c$. 
This relies on a representation of the final state $|\Phi_f\rangle$ that has $N+1$ valence electrons (the extra electron is the photoelectron). 
We extend the interpretation of the Kohn-Sham orbitals in the core hole approach and assume that a single Slater determinant of arbitrary $N+1$ filled \emph{final-state} Kohn-Sham orbitals is a good approximation of $|\Phi_f\rangle$, an eigenstate of the core-excited Hamiltonian.
Now the final-state index $f$ refers to a configuration of final-state orbitals $f=(f_1, f_2, \cdots, f_{N+1})$ that the $N+1$ electrons can occupy.
Furthermore, the energy of $|\Phi_f\rangle$ (relative to the threshold energy $E_\text{th}$) is approximated as a non-interacting summation of the energies of all occupied final-state orbitals: $E_f=\sum^{N}_{\mu=1}\tilde{\varepsilon}_{f_\mu}-E_\text{th}$, where the threshold energy $E_\text{th}=\min_f E_f$, corresponds to the first final state configuration $(1,2,...,N+1)$.
$|\Phi_f\rangle$ can be represented as 
\begin{align}
\begin{split}
&|\Phi_f\rangle=\prod_{\mu=1}^{N + 1} \tilde{\phi}^\dagger_{f_{\mu}}  \prod_{\nu=1}^{N}b^\dagger_{\nu}  h^\dagger |\Phi_i\rangle \\
&=\prod_{\mu=1}^{N+1}  (\sum^{\text{empty}}_{c} \xi_{f_{\mu},c} a^\dagger_c +\sum^{\text{occupied}}_{v}\xi_{f_{\mu}, v} b_v)  \prod_{\nu=1}^{N}b^\dagger_{\nu}   h^\dagger |\Phi_i\rangle \\
\end{split}
\label{eq:many-body}
\end{align}
in which all the $N+1$ electrons are first removed via $h^\dagger$ and $N$ $b^\dagger_v$'s from the initial state, and then recreated via $N+1$ $\tilde{\phi}^\dagger_{f_\mu}$'s to define the many-body final state. 
Note that a final-state orbital $\phi_f$ can in general be a hybridization of both the occupied ($v$) and empty ($c$) orbitals in the initial state due to the core-hole perturbation. $\xi_{ij}$'s are the transformation coefficients from the initial to final state \cite{sm}. 
As the core-hole effect on each initial-state orbital accumulates, $|\Phi_f\rangle$ in Eq. (\ref{eq:many-body}) reproduces the full many-electron response to the core-hole potential [Eq. (\ref{eq:Phi_f})].
Ultimately, the expression for $A^f_c$ can be obtained by combining the coefficients of the $a^\dagger_c h^\dagger$ term
\begin{align}
\begin{split}
A^f_c =\det
\begin{bmatrix}
\xi_{f_1, 1} & \xi_{f_1, 2} & \cdots & \xi_{f_1, N} & \xi_{f_1, c} \\
\xi_{f_2, 1} & \xi_{f_2, 2} & \cdots & \xi_{f_2, N} & \xi_{f_2, c} \\
\vdots & & \ddots & & \vdots\\
\xi_{f_{N+1}, 1} & \xi_{f_{N+1}, 2} & \cdots & \xi_{f_{N+1}, N} & \xi_{f_{N+1}, c} \\
\end{bmatrix}
\label{eq:Afc}
\end{split}
\end{align}
Similar determinant expressions were also obtained in previous work \cite{anderson1967infrared, stern1983many, ohtaka1990theory} but they are rarely applied in a solid-state context from first-principles. Thus it is of great interesting to examine whether the derived many-body $\Delta$SCF formalism in Eq. (\ref{eq:many-body}) and (\ref{eq:Afc}) can reproduce the correct lineshapes for the investigated TMOs.

\begin{figure}[t]
  \centering
  \includegraphics[angle=0, width=0.99\linewidth]{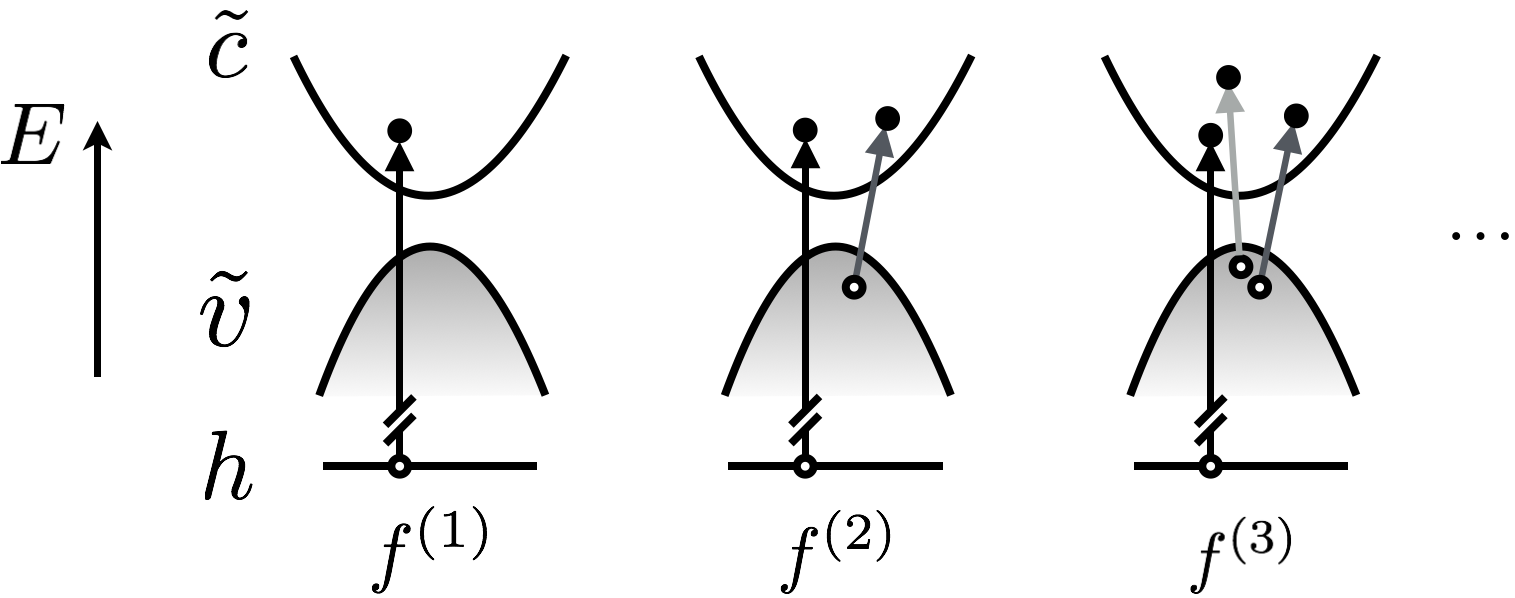}
  \caption{Schematic of the excitation configurations in the final-state space.  $h$ indicates the single core level while $\tilde{c}$ and $\tilde{v}$ indicate the empty and occupied orbtial spaces respectively. }
  \label{fig:schematics}
\end{figure}

Evaluating $A^f_c$ seems formidable at first glance because, unlike small molecules, a solid contains many electrons, which leads to huge number combinations of final states.
To appreciate this, we regroup the final-state configurations according to the number of \ehpair{} pairs excited, by analogy to the initial-state decomposition in Eq. (\ref{eq:Phi_f}).
For example, we denote a single configuration with one core-excited \ehpair{} pair as $f^{(1)}=(1,2,\cdots,N, f_{N+1})$, which is typically used in producing the one-body XAS in the core-hole approach and $f_{N+1}$ is the orbital the photoelectron will occupy in the final state. Based on this configuration, we can define a double configuration with one more valence \ehpair{} pair as $f^{(2)}=(1,\cdots, i-1, i+1, \cdots, N, f_{N}, f_{N+1})$, where $f_{N+1}>f_{N} >N$, and so forth (Fig. \ref{fig:schematics}).
Despite a large number of excitation configurations, not all of them contribute equally to the XANES absorbance.
First, for an insulator, the energy $\Omega_f$ of the final states will increase rapidly with the number of \ehpair{} pairs excited across the band gap ($E_\text{g}$): $E_f \geq (n-1)E_\text{g}$.
This largely reduces the number of necessary final-states for producing the first few eVs of XANES for an insulator. 
Second, the many-body overlap may decrease rapidly with increasing number of \ehpair{} pairs excited, as we will show next.

\begin{figure}[t]
  \centering
  \includegraphics[angle=0, width=0.99\linewidth]{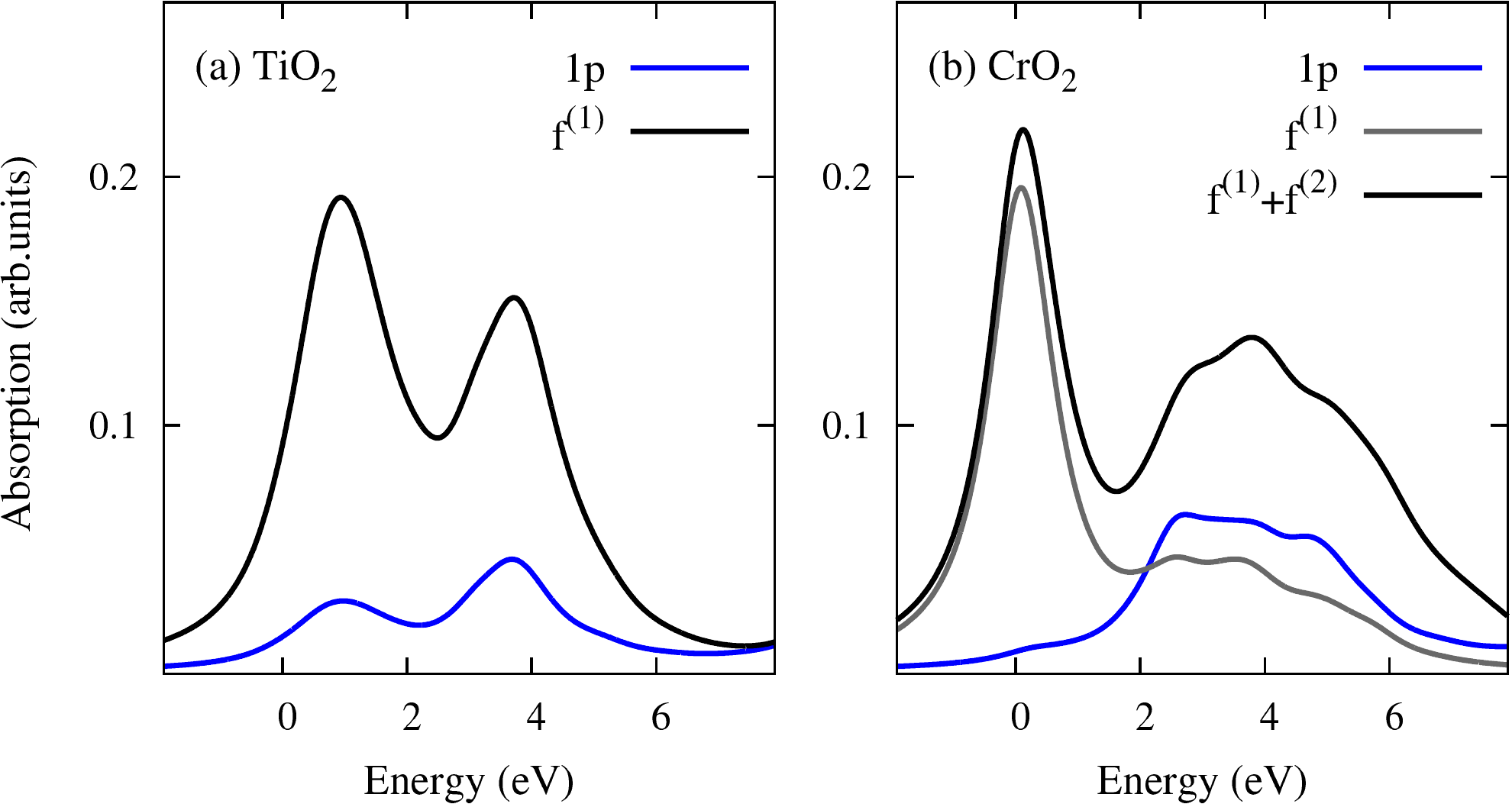}
  \caption{$\Delta$SCF XAS rectified by the many-body matrix formalism for \ce{TiO2} (a) and \ce{CrO2} (b). For comparison, the spectra produced by the one-body FCH formalism are denoted as 1p FCH. The absorption threshold is aligned at $E=0$. 
  }
  \label{fig:det_xas}
\end{figure}

Fig. \ref{fig:det_xas} shows the XAS calculated from the many-body matrix element as in Eq. (\ref{eq:many-body}) and (\ref{eq:Afc}). The wavefunctions and the overlap matrix elements $\xi_{ij}$ used for $A^f_c$ are obtained from exactly the same set of FCH calculations \cite{sm} as previously described. For simplicity, only the states at the $\Gamma$-point of the supercell Brillouin zone are used for producing the XAS. We reexamine two extreme cases: the insulating \ce{TiO2} and the metallic \ce{CrO2}. Remarkably, the simulated XAS lineshapes with the many-body correction are in excellent agreement with experiments. In particular for \ce{CrO2},  the edge sharpness is completely retrieved despite the disappearance of the first peak in a single-body theory.

When calculating the XAS for \ce{TiO2}, we find its spectrum already converges at the order of $f^{(1)}$. As expected, the contributions from $f^{(2)}$ are at higher energies due to the sizable band gap, and are much weaker due to reduced many-body overlap. However, it is harder to achieve convergence in the metallic \ce{CrO2}. While the first peak is retrieved mainly from low-order excitations $f^{(1)}$, the correct peak-intensity ratio can only be reproduced when the next-order $f^{(2)}$ is taken into account, which substantially intensifies the absorption feature, $\sim4$ eV above onset. We find $f^{(3)}$ can be neglected up to the first $8$ eV.

At last, we explain why the one-body $\Delta$SCF formalism tends to underestimate the peak intensity. With S constant, Eq. (\ref{eq:matrix_element}) produces nothing but an RPA absorption spectrum in the final-state picture in which the charge density has equilibrated with the core-hole potential. No charge relaxation dynamics are involved. In the this adiabatic final state of \ce{O} $1s$ excitations, a substantial amount of electron density is transferred onto the $2p$ state of the excited \ce{O} ($0.55$ $e^-$ for \ce{TiO2} and $0.72$ $e^-$ for \ce{CrO2}), which eventually blocks some bright transitions from $1s\rightarrow2p$ and leads to the reduced pre-edge intensity. On the other hand, both BSE and the many-body $\Delta$SCF formalism have taken into account the excitation dynamics from the initial state to the final state. In both the BSE [Eq. (\ref{eq:tda})] and the many-body formalism [Eq. (\ref{eq:many-body})], only the initial empty states $\phi_c$ are involved and thus the edge sharpness is captured.

\paragraph{Conclusions} We have shown that the BSE and a newly developed many-body $\Delta$SCF approach are highly predictive for \ce{O} $K$-edge fingerprints of TMOs. The systematic underestimation of the peak-intensity ratio within the original one-body $\Delta$SCF approach is attributed to the absence of many-electron excitation dynamics in this formalism. We have demonstrated how to rectify these shortcomings (1) an expansion of the final-state configurations in terms of Slater determinants; (2) a projection of these final states onto the initial-state orbital basis; and (3) a correct accounting of the relevant component of the adiabatic final state accessible to dipole transitions. This many-body formalism is transferable and not at all not peculiar to TMOs. In fact, the expansion in Eq. (\ref{eq:many-body}) goes beyond the BSE formalism and is more universal. We leave the discussion of shakeup effects, improvement of efficiency, and other examples to near-future work. 

\paragraph{Acknowledgement} Theoretical and computational work was performed by Y. L. and D. P. at The Molecular Foundry, which is supported by the Office of Science, Office of Basic Energy Sciences, of the United States Department of Energy under Contact No. DE-AC02-05CH11231. We acknowledge fruitful discussion with Chunjing Jia (Y. L.) and Bill Gadzuk (J. V.). Computations were performed with the
computing resources at the National Energy Research Scientific Computing Center (NERSC).


\end{document}